\pdfoutput=1
\documentclass[twocolumn,10pt,aps,prd,preprintnumbers,amsmath,amssymb,superscriptaddress,nofootinbib]{revtex4-1}

\usepackage{amssymb,amsmath,bbold}
\usepackage{graphicx,soul}
\usepackage[usenames,dvipsnames]{xcolor}
\usepackage[unicode=true,pdfusetitle,
 bookmarks=true,bookmarksnumbered=false,bookmarksopen=false,citecolor=Turquoise,
 breaklinks=false,pdfborder={0 0 1},backref=false,colorlinks=true,pdfpagemode=FullScreen]
 {hyperref}

\usepackage{tikz}

\usetikzlibrary{arrows,calc,graphs,patterns,positioning}
\usetikzlibrary{decorations,decorations.markings,decorations.pathmorphing,decorations.pathreplacing}
\usetikzlibrary{shapes, shapes.geometric}

\tikzset{/pgf/decoration/.cd,
    number of sines/.initial=5,
    angle step/.initial=20,
}
\newdimen\tmpdimen

\pgfdeclaredecoration{complete sines}{initial}
{
    \state{initial}[
        width=+0pt,
        next state=move,
        persistent precomputation={
            \pgfmathparse{\pgfkeysvalueof{/pgf/decoration/angle step}}%
            \let\anglestep=\pgfmathresult%
            \let\currentangle=\pgfmathresult%
            \pgfmathsetlengthmacro{\pointsperanglestep}%
                {(\pgfdecoratedremainingdistance/\pgfkeysvalueof{/pgf/decoration/number of sines})/360*\anglestep}%
        }] {}
    \state{move}[width=+\pointsperanglestep, next state=draw]{
        \pgfpathmoveto{\pgfpointorigin}
    }
    \state{draw}[width=+\pointsperanglestep, switch if less than=1.25*\pointsperanglestep to final, 
        persistent postcomputation={
        \pgfmathparse{mod(\currentangle+\anglestep, 360)}%
        \let\currentangle=\pgfmathresult%
    }]{%
        \pgfmathsin{+\currentangle}%
        \tmpdimen=\pgfdecorationsegmentamplitude%
        \tmpdimen=\pgfmathresult\tmpdimen%
        \divide\tmpdimen by2\relax%
        \pgfpathlineto{\pgfqpoint{0pt}{\tmpdimen}}%
    }
    \state{final}{
        \ifdim\pgfdecoratedremainingdistance>0pt\relax
            \pgfpathlineto{\pgfpointdecoratedpathlast}
        \fi
   }
}

\tikzset{
    scalar/.style={draw=black, thick, dashed},
    fermion/.style={draw=black, postaction={decorate}, decoration={markings, mark=at position .55 with {\arrow[]{latex}}}, thick},
    antifermion/.style={draw=black, postaction={decorate}, decoration={markings, mark=at position .55 with {\arrow{latex}}}, thick},
    photon/.style={draw=none, outer ysep=3pt, postaction={draw=black, thick, decorate}, decoration={complete sines, amplitude=7pt}},
    photonloop/.style={draw=none, outer ysep=3pt, postaction={draw=black, thick, decorate}, decoration={complete sines, number of sines=15, amplitude=5pt}},
    gluon/.style={draw=none, outer ysep=3pt, postaction={draw=black, thick, decorate}, decoration={coil, amplitude=3pt, segment length=5pt}}, 
    composite/.style={draw=gray!70!, line width=5pt},
    vtx/.style={inner sep=0pt},
    mass/.style={cross out, draw, minimum size=5pt, inner sep=0pt},
    ins/.style={draw, cross, shape=circle, minimum size=5pt, inner sep=0pt}, 
    cross/.style={path picture={\draw[black] (path picture bounding box.south east) -- (path picture bounding box.north west) (path picture bounding box.south west) -- (path picture bounding box.north east);}},
    vertex/.style={draw, shape=circle, fill=black, minimum size=3pt, inner sep=0pt},
    blob/.style={draw, shape=circle, preaction={fill,white}, pattern = north west lines, minimum size=15pt, inner sep=0pt},        
    loop/.style={shape=circle, minimum size=1.7cm, outer sep=9pt},
    site/.style={draw, shape=circle, minimum size=1.5cm, inner sep=0pt},
    brane/.style={trapezium, draw, trapezium stretches=true, minimum height=4cm, minimum width=3.5cm, inner sep=0, trapezium left angle=35, trapezium right angle=145, shape border uses incircle, shape border rotate=17.5},
    momentum/.style={->, dist/.store in=\segDistance, pos/.store in=\segPos, len/.store in=\segLength,
      to path={
      ($(\tikztostart)!\segPos!(\tikztotarget)!\segLength/2!(\tikztostart)!\segDistance!90:(\tikztotarget)$) -- 
      ($(\tikztostart)!\segPos!(\tikztotarget)!\segLength/2!(\tikztotarget)!\segDistance!-90:(\tikztostart)$)  \tikztonodes
      }, 
      pos=.5,
      len=7mm,
      dist=2mm
    },    
}

\begin{document}

\preprint{IPMU17-0098}

\title{LHC Search for Right-handed Neutrinos in $Z^\prime$ Models}

\author{Peter Cox}
\email[]{peter.cox@ipmu.jp}
\affiliation{Kavli IPMU (WPI), UTIAS, University of Tokyo, Kashiwa, Chiba 277-8583, Japan}
\author{Chengcheng Han}
\email[]{chengcheng.han@ipmu.jp }
\affiliation{Kavli IPMU (WPI), UTIAS, University of Tokyo, Kashiwa, Chiba 277-8583, Japan}
\author{Tsutomu T. Yanagida}
\email[]{tsutomu.tyanagida@ipmu.jp}
\affiliation{Kavli IPMU (WPI), UTIAS, University of Tokyo, Kashiwa, Chiba 277-8583, Japan}
\affiliation{Hamamatsu Professor}

\begin{abstract}
We consider right-handed neutrino pair production in generic $Z^\prime$ models. We propose a new, model-independent analysis using final states containing a pair of same-sign muons. A key aspect of this analysis is the reconstruction of the RH neutrino mass, which leads to a significantly improved sensitivity. Within the $U(1)_{(B-L)_{3}}$ model, we find that at the HL-LHC it will be possible to probe RH neutrino masses in the range $0.2\lesssim M_{N_R} \lesssim 1.1\,$TeV.
\end{abstract}

\maketitle
\section{Introduction}

The observation that neutrinos have non-zero masses requires the existence of physics beyond the Standard Model (SM). 
One of the simplest and most well-motivated SM extensions is the inclusion of right-handed (RH) neutrinos; the observed neutrino masses and mixings can then be explained via the Type-I see-saw mechanism~\cite{Minkowski:1977sc, *Yanagida:1979as, *Glashow:1979nm, *GellMann:1980vs}. 
While the small observed neutrino masses may hint at very large Majorana masses for the RH neutrinos, this need not necessarily be the case; indeed, small Yukawa couplings appear to be prevalent amongst the charged leptons.
It is therefore interesting to consider the possibility that (at least some of) the RH neutrinos may be relatively light and hence be directly accessible at LHC energies.

There have already been several searches for TeV scale RH neutrinos at the LHC~\cite{1501.05566, 1506.06020, 1603.02248, CMS-PAS-EXO-16-045}. 
Within the minimal model, the RH neutrinos are produced via mixing with the light neutrinos (see eg.~\cite{0901.3589, 1502.06541}), which generally leads to small production cross-sections. 
Although large mixing is also possible~\cite{0705.3221, 0907.1607}, current LHC searches are generally not competitive with other constraints~\cite{1702.04668}.  
However, in many well-motivated SM extensions there can be new production mechanisms, often via the decay of a new heavy resonance. 
This is the case for example in left-right symmetric models~\cite{Keung:1983uu}, or a gauged $B-L$ model where the RH neutrinos can be produced through the decay of the new $Z^\prime$ gauge boson~\cite{Buchmuller:1991tu}. 
In this paper, we investigate the potential of (HL-)LHC searches for RH neutrino pair production via the decay of a general $Z^\prime$ gauge boson. 

The Lagrangian in the neutrino sector is given by
\begin{equation}
  \mathcal{L}_{Y}= - \bar{l}_L Y \tilde{H} N_R -\frac{1}{2} \bar{N}_R M_N N_R^c+ h.c. \,,
\end{equation}
leading to three main decay modes for the RH neutrinos:
\begin{align}
  &\Gamma(N_i \rightarrow l_j^\pm W^\mp)=\frac{g^2 m_N^3}{64\pi m_W^2} |(U_{\ell N})_{ji}|^2(1-\frac{3 m_W^4}{m_N^4}+\frac{2m_W^6}{m_N^6}),  \notag \\
  &\Gamma(N_i \rightarrow \nu Z)=\frac{g^2 m_N^3}{64\pi m_W^2} \sum_j|(U_{\ell N})_{ji}|^2(1-\frac{3 m_Z^4}{m_N^4}+\frac{2m_Z^6}{m_N^6}),  \notag \\
  &\Gamma(N_i \rightarrow \nu h)=\frac{g^2 m_N^3}{64\pi m_W^2} \sum_j|(U_{\ell N})_{ji}|^2(1-\frac{ m_h^2}{m_N^2})^2,
\end{align}
where $U_{\ell N}=U_{PMNS}\, m_\nu^{1/2} \Omega M_N^{-1/2}$, with $\Omega$ an arbitrary complex orthogonal matrix\footnote{For a detailed discussion of the neutrino mixing see eg. ~\cite{0901.3589}.}. 
The lifetimes of the heavy neutrinos are then largely determined by their mass, and the mass scale of the light active neutrinos.
For $m_N\lesssim200\,$GeV, they can have macroscopic decay lengths~\cite{0812.4313, 0907.4186} and are best searched for using displaced vertices~\cite{0907.4186, 1604.06099, 1609.03594, 1610.03894}. Here, we focus on the alternative case, $m_N\gtrsim200\,$GeV.

The most promising search channel involves RH neutrino decays to charged leptons, since these give rise to clean same-sign di-lepton final states, with relatively small SM backgrounds. 
Preliminary studies of the di-lepton~\cite{0907.4186}, tri-lepton~\cite{0812.4313} and four lepton~\cite{0803.2799} channels were performed in the context of the $B-L$ model, prior to the start of LHC running. 
More recently, the prospects for tri-lepton and boosted Higgs final states were investigated in~\cite{1512.08373}, while searches based on neutrino jets have been proposed for when the $N_R$ is highly boosted~\cite{1607.03504, 1610.08985}. 
Furthermore, current same-sign di-lepton (and tri-lepton) searches~\cite{1412.0237, 1501.05566, 1506.06020, 1603.02248, CMS-PAS-EXO-16-045} now have limited sensitivity to RH neutrino production, although limits are rarely presented in the context of $Z^\prime$ models. 

In this work, we aim to highlight the benefits of a dedicated analysis. 
In particular, we demonstrate that with increased integrated luminosity it will become possible to reliably reconstruct the RH neutrino mass. 
Beyond providing valuable information about the mass in the event of a discovery, we show that this approach also leads to significantly improved sensitivity.

\section{Same-sign Muon Search} 

\begin{figure}[ht]
  \resizebox{7cm}{!}{
  \begin{tikzpicture}
  \large
  \node[vtx, label=180:$q$] (i1) at (0,1.4) {};
  \node[vtx, label=180:$\bar{q}$] (i2) at (0,-1.4) {};
  \coordinate[] (v1) at (1.6,0) {};
  \coordinate[] (v2) at (4.1,0) {};
  \coordinate[] (v3) at (5.7,1.4) {};
  \coordinate[] (v4) at (5.7,-1.4) {};
  \node[vtx, label=0:$\mu^-$] (o1) at (7.4,2.2) {};
  \node[vtx, label=0:$W^+$] (o2) at (7.4,0.5) {};
  \node[vtx, label=0:$W^+$] (o3) at (7.4,-0.5) {};
  \node[vtx, label=0:$\mu^-$] (o4) at (7.4,-2.2) {};
  \graph[use existing nodes]{
     i1 --[fermion] v1 --[fermion] i2;
     v1 --[photon, edge label'=$Z^{\prime(*)}$] v2;
     v2 --[style={draw=black, thick}, edge label=$N_R$]v3;
     v2 --[style={draw=black, thick}, edge label'=$N_R$] v4;
     o2 --[photon] v3 --[fermion] o1;
     o3 --[photon] v4 --[fermion] o4;
  };
  \end{tikzpicture}
  }
  \caption{RH neutrino pair production via an s-channel $Z^\prime$.}
  \label{fig:feynman}
\end{figure}
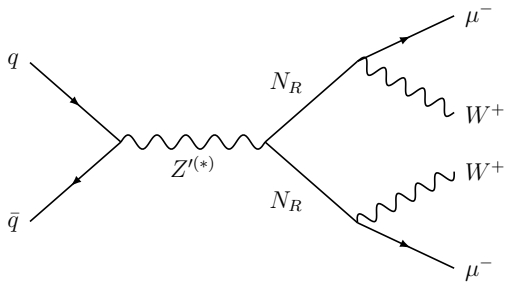

In this section we propose in detail a model-independent search for RH neutrinos pair produced via the decay of a new heavy $Z^\prime$ resonance. 
We restrict our focus to the same-sign di-muon channel, due to increased backgrounds for electron final states that require a data-driven approach to reliably estimate. 
However, a similar search strategy could in principle be applied to $e^\pm e^\pm$ and $e^\pm \mu^\pm$ final states. 
The relevant process is shown in Fig.~\ref{fig:feynman}.

\subsection{Benchmark model}

Although our search is model-independent, we adopt a benchmark model in order to clearly demonstrate the future sensitivity. 
Perhaps one of the most commonly considered $Z^\prime$ models is gauged $B-L$, where the presence of three RH neutrinos can be further motivated by anomaly cancellation. 
However, this model is already highly constrained by $Z^\prime\to\mu^+\mu^-$ resonance searches. 
Regions of parameter space where one has sensitivity to $N_R$ production, even at the HL-LHC, are therefore either already excluded or likely to be excluded in the near future.  

We shall instead consider a related model, $U(1)_{(B-L)_{3}}$. 
This is a flavoured $B-L$ gauge symmetry under which only the third generation fermions are charged. 
In this case anomaly cancellation requires a single RH neutrino\footnote{For simplicity, we assume $N_R$ decays dominantly to second generation leptons; the required coupling can be generated by $U(1)_{(B-L)_3}$ breaking (see Ref.~\cite{1705.03858}).}, whose Majorana mass is naturally of order the $U(1)_{(B-L)_3}$ breaking scale. 
Further details of the model can be found in Ref.~\cite{1705.03858}.
Interestingly, this model admits the possibility that two additional RH neutrinos could have super-heavy Majorana masses and generate the observed baryon asymmetry via leptogenesis~\cite{Fukugita:1986hr, hep-ph/0208157}. 
It was also recently considered as an explanation for certain anomalies observed in rare $B$ decays~\cite{1704.08158, 1705.03858, 1705.00915}. 

The coupling of the $Z^\prime$ to the SM fermions in this model is, in the gauge basis,
\begin{equation}
  \mathcal{L}_{Z^\prime}= g Z_\mu^\prime(\frac13\bar{t}\gamma^\mu t +  \frac13\bar{b}\gamma^\mu b - \bar{\tau}\gamma^\mu \tau- \bar{\nu_L}\gamma^\mu \nu_L -\bar{N}_R\gamma^\mu N_R) \,.
\end{equation}
We assume a gauge coupling $g=0.6$ for all our benchmark points. 
After $U(1)_{(B-L)_3}$ breaking and rotation to the mass basis, there may also be couplings to the first and second generation fermions. 
These will not play a significant role in what follows and we assume them to be small. 
The dominant $Z^\prime$ production mechanism at the LHC is then $b\bar{b}\to Z^\prime$. Direct searches for the $Z^\prime$ in the $\tau\tau$ final state~\cite{1608.00890, 1611.06594} currently impose $M_{Z^\prime}\gtrsim900\,$GeV for $g=0.6$. 
Searches using the $t\bar{t}$ and $b\bar{b}$ final states do not currently provide competitive bounds. 

\subsection{SM Backgrounds}

The production of same-sign leptons is a relatively rare process within the SM and the background can be divided into two main classes. 
Firstly, there is the prompt background originating from decays of $W$, $Z$ and $t$, with the dominant processes being $WZ$ and $\bar{t}tW/\bar{t}tZ$ production. 
Secondly, there are non-prompt leptons produced via the decays of long-lived particles, predominantly the semi-leptonic decays of heavy flavour hadrons. 
In existing searches this background is determined using a data-driven approach. 
For large di-lepton invariant masses, which will be relevant to our analysis, the non-prompt contribution to the total background is found to be less than 15\% in existing analyses~\cite{1412.0237} and will therefore be neglected in the following. 
Lastly, there can be additional sources of background due to lepton charge misidentification and jets misidentified as leptons. 
These are important for searches involving electrons, but are negligible in the di-muon channel.

In order to estimate the prompt background, we generate monte carlo event samples containing same-sign muon pairs from $WZ$ and $\bar{t}tW/\bar{t}tZ$ production with {\tt MadGraph-2.5.4}~\cite{1405.0301} and {\tt PYTHIA-6.4}~\cite{hep-ph/0603175}, followed by detector simulation with {\tt Delphes 3}~\cite{1307.6346}. 
These include muons arising both directly from $W/Z/t$ decays and also from subsequent decays of tau leptons. 
A weighted approach is used, with events generated in bins of $H_T$ in order to ensure sufficient statistics for studies at high integrated luminosity. 
In the case of $WZ$ production, events are generated with up to two additional hard jets using MLM matching~\cite{hep-ph/0602031}. 
The samples are generated at leading order and then normalised to the latest calculations of the total production cross-section for $WZ$ at NNLO~\cite{1604.08576} and $\bar{t}tW/\bar{t}tZ$ at NLO~\cite{1610.07922}.

\subsection{Initial Selection} \label{sec:initial_selection}

The initial event selection closely resembles that used in the ATLAS 8 TeV same-sign lepton search~\cite{1412.0237}. 
We require at least one pair of isolated, same-sign muons satisfying $p_T>25,\,20\,$GeV and $|\eta|<2.4$. 
The di-muon invariant mass is also required to satisfy $M_{\mu\mu}>15\,$GeV. 
We allow for the presence of additional leptons in the event, since this significantly increases the number of signal events by providing sensitivity to both the two and three muon final states, where up to one of the $W$ bosons decays leptonically. 
Events containing a pair of opposite-sign, same-flavour leptons consistent with the $Z$ mass, $|M_{ll}-m_Z|<10\,$GeV, are vetoed in order to reduce the SM background.

The novel aspect of this analysis is the ability to reconstruct the right-handed neutrino mass. 
In addition to the above selections, we therefore require at least one hadronically decaying $W$ boson candidate. 
Depending on the $Z^\prime$ mass, the $W$ bosons from $N_R$ decays may be sufficiently boosted to exploit jet substructure methods for W boson tagging. 
On the other hand, for lighter $Z'$ masses the two jets from the $W$ decay will be resolved separately. 
We allow for both possibilities in our analysis and reconstruct up to two hadronically decaying W bosons as described below.

Firstly, we identify hadronically decaying boosted $W$ bosons following the procedure adopted by CMS for their 13 TeV analyses (eg.~\cite{CMS-PAS-B2G-17-001}). 
We begin with jets clustered using the anti-$k_T$ algorithm~\cite{0802.1189}, as implemented in {\tt FastJet}~\cite{1111.6097}, with a distance parameter $R=0.8$. 
These jets are further required to satisfy $p_T>200\,$GeV and $|\eta|<2.5$. 
Jet grooming is then performed using the soft-drop algorithm~\cite{1402.2657} with $\beta=0$ and $z_{\text{cut}}=0.1$. 
For the jet to be considered as a $W$ boson, the soft-drop jet mass is required to satisfy $65<M_{jet}<95\,$GeV. 
Finally, the N-subjetiness~\cite{1011.2268} ratio $\tau_{21}=\tau_2/\tau_1$ should satisfy $\tau_{21}<0.75$, which provides further discrimination against gluon and single-quark initiated jets while maintaining $\sim100\%$ signal efficiency for large $Z^\prime$ masses.

For events containing fewer than two reconstructed boosted $W$ bosons, we also reconstruct $W$ bosons decaying into two resolved jets. 
Jets are clustered using the anti-$k_T$ algorithm with distance parameter $R=0.5$ and are required to satisfy $p_T>25\,$GeV and $|\eta|<2.4$. 
If the event contains a boosted $W$ boson, the jets are additionally required to satisfy $\Delta R(W,\text{jet})>0.8$. $W$ boson candidates are then reconstructed iteratively by choosing the pair of jets which minimises $|M_{jj}-m_W|$. 
Finally, the di-jet mass is required to satisfy $50<M_{jj}<110\,$GeV in order to be identified as a $W$ boson.

\subsection{$N_R$ Reconstruction}

\begin{figure}[ht]
  \centering
  \includegraphics[width=0.5\textwidth]{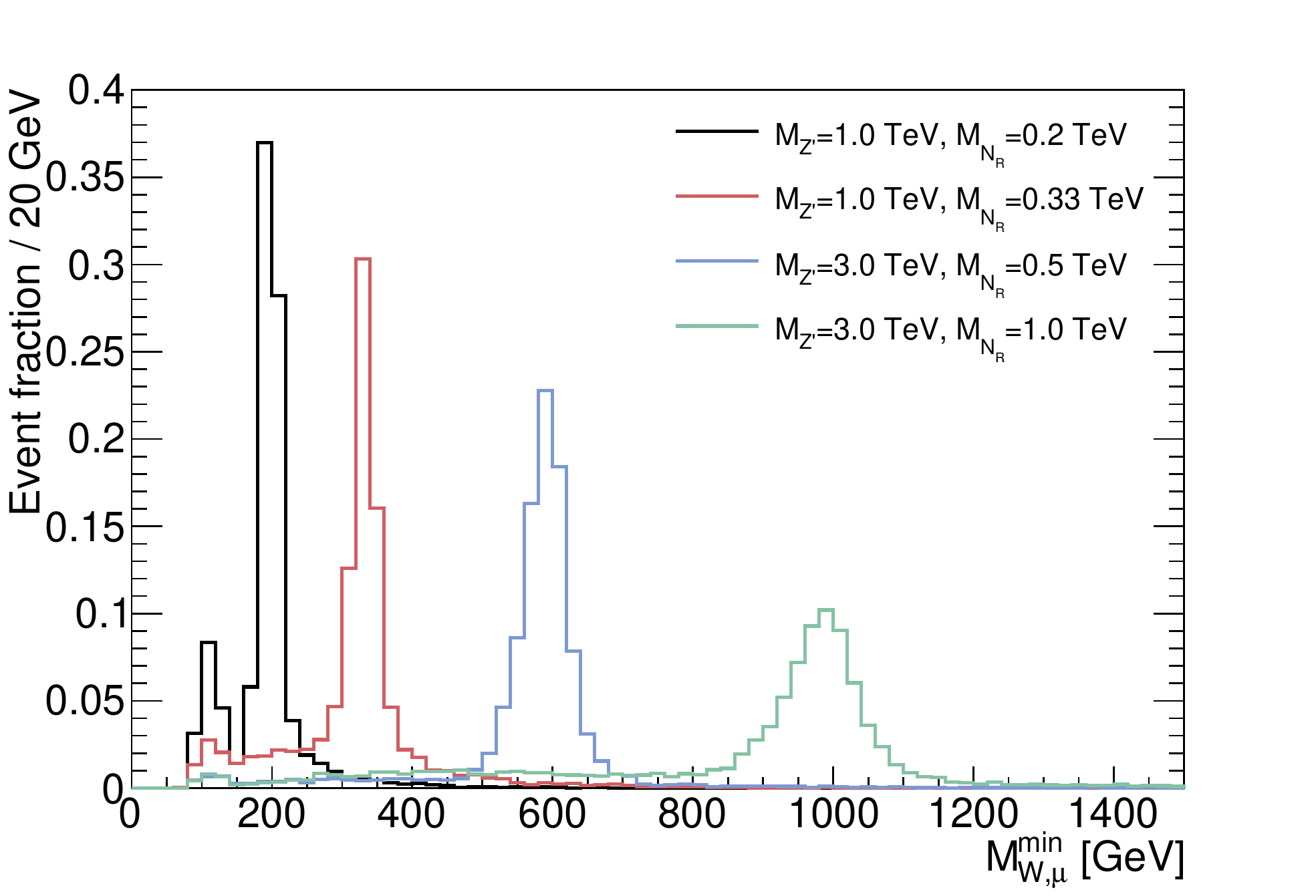} 
  \includegraphics[width=0.5\textwidth]{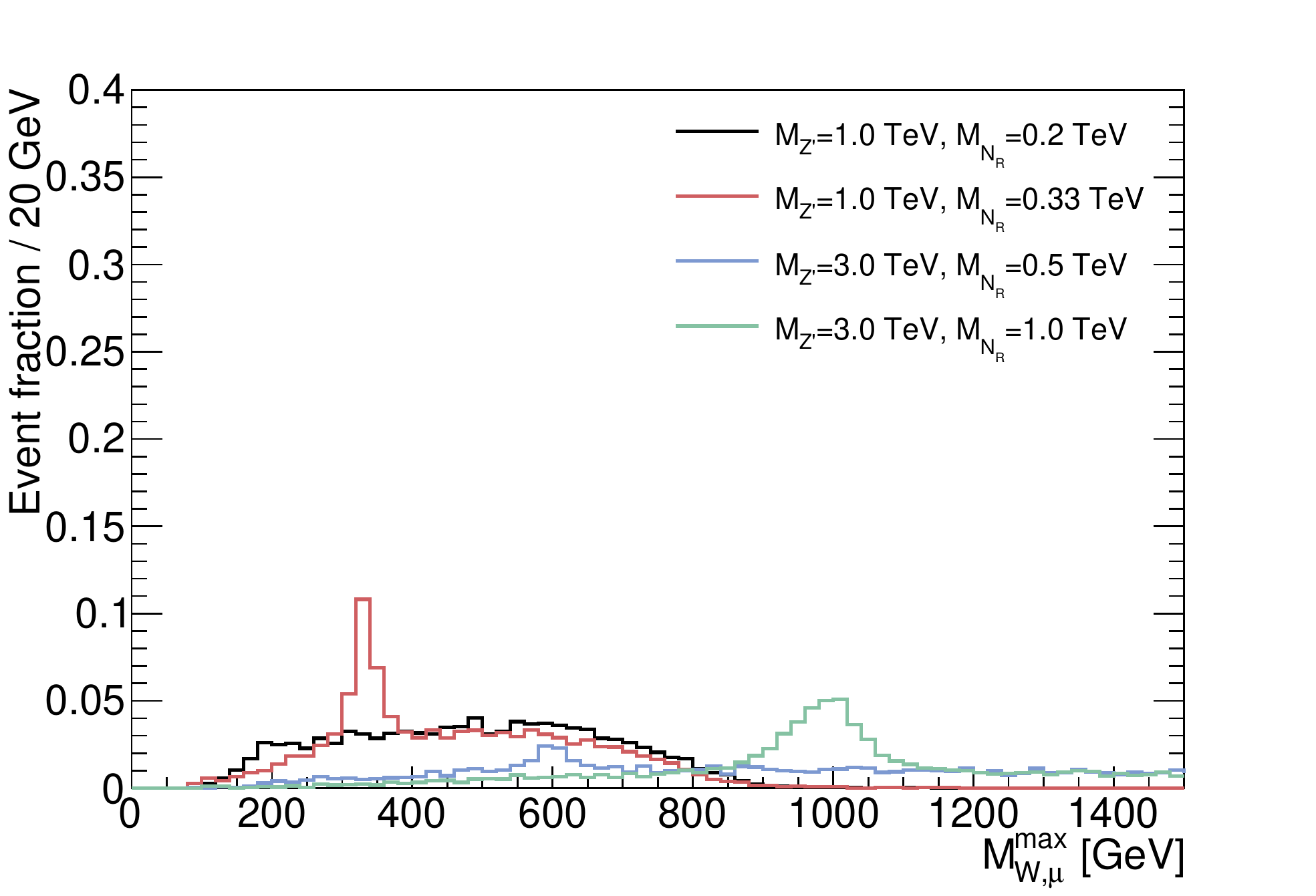} 
  \caption{The ($W,\mu$) invariant mass obtained in $1W$ events using the muon with the smallest (top) and largest (bottom) $\delta R$ separation.}
  \label{fig:mNR_1W}
\end{figure}

\begin{figure}[ht]
  \centering
  \includegraphics[width=0.5\textwidth]{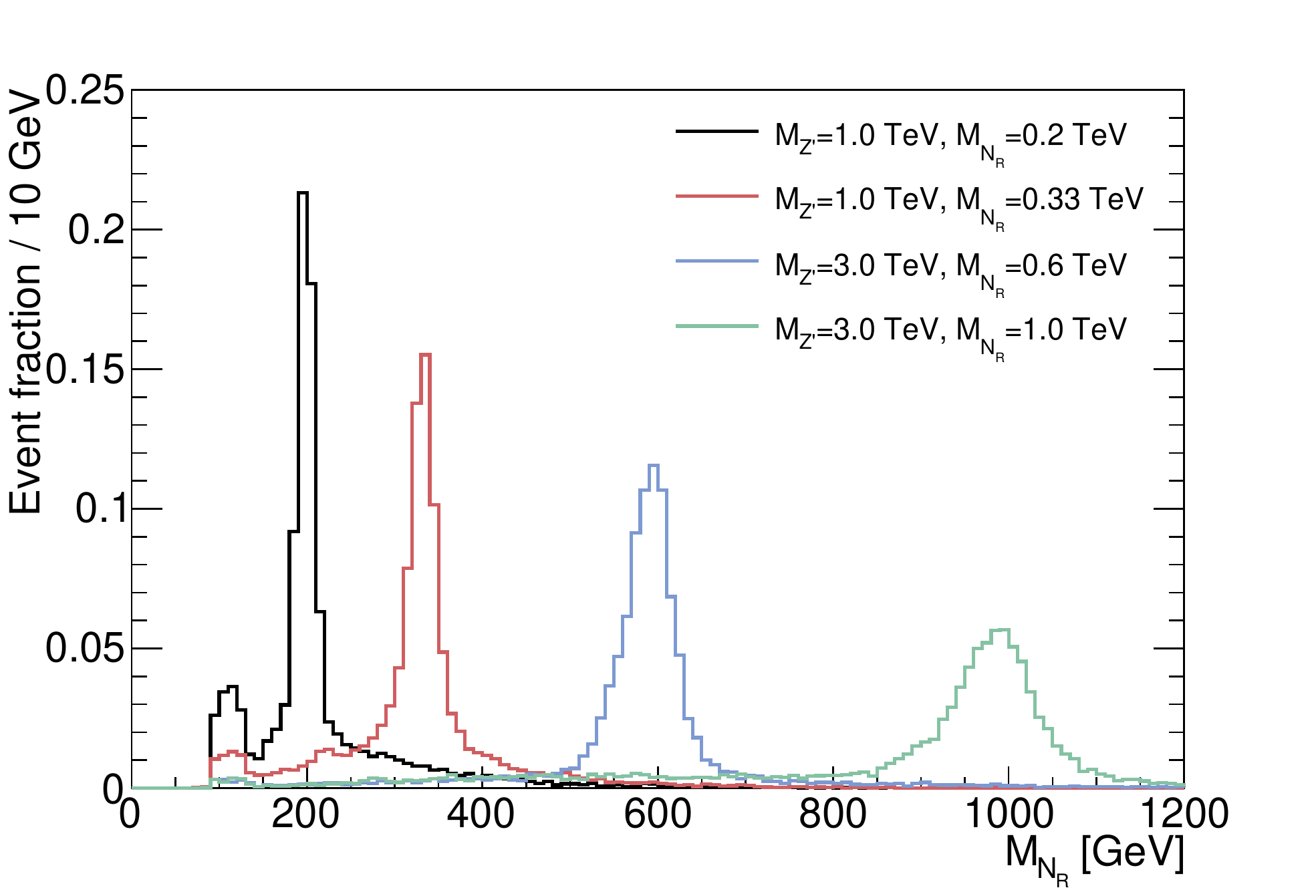} 
  \caption{Reconstructed $N_R$ mass for several benchmark signals.}
  \label{fig:mNR}
\end{figure}

In reconstructing the $N_R$ mass we are faced with two choices when selecting the $W$ boson and muon to associate with each $N_R$ decay. 
For events containing two reconstructed $W$ bosons, the solution is straightforward: requiring consistency of the two $N_R$ masses, one chooses the combination which minimises the difference in the reconstructed masses $|M_{N_1}-M_{N_2}|$, where $M_{N_{1,2}}$ are the invariant masses of the two $(W,\mu)$ systems. 
The final reconstructed $N_R$ is taken as the mean of $M_{N_1}$ and $M_{N_2}$ in the case of two boosted or two resolved $W$ bosons. 
In events containing one boosted and one resolved $W$, only the boosted $W$ is used to determine the final $N_R$ mass as this leads to slightly improved mass resolution.

The situation is somewhat more complicated in events with only a single reconstructed $W$ boson. 
In this case it is, a priori, not clear which muon to identify as originating from the same $N_R$ decay as the $W$. 
However, in practice, simply selecting the muon with the smallest $\delta R$ separation from the $W$ allows one to reliably reconstruct the true $N_R$ mass. 
This can be clearly seen in Fig.~\ref{fig:mNR_1W}, where we show the reconstructed $N_R$ masses obtained using the two possible combinations. 
As is to be expected, this approach performs better when $M_{N_R}\ll M_{Z^\prime}$ due to the higher boost of the $N_R$ decay products. 

The final reconstructed $N_R$ mass (including both $1W$ and $2W$ events) is shown in Fig.~\ref{fig:mNR}. 
We find that it's possible to obtain good mass resolution, of the order of $\sim10\%$, depending on the $N_R$ and $Z^\prime$ masses. 
This mass resolution will ultimately determine the $M_{N_R}$ bin width in any future search. 
Here, we simply require instead that
\begin{equation} \label{eq:masscut}
  |M_{N_R}^\text{reco}-M_{N_R}|<0.1M_{N_R} \,,
\end{equation}
where $M_{N_R}$ is the ``true" right-handed neutrino mass for each benchmark signal point.

\subsection{Final Selection}

\begin{figure}[ht]
  \centering
  \includegraphics[width=0.5\textwidth]{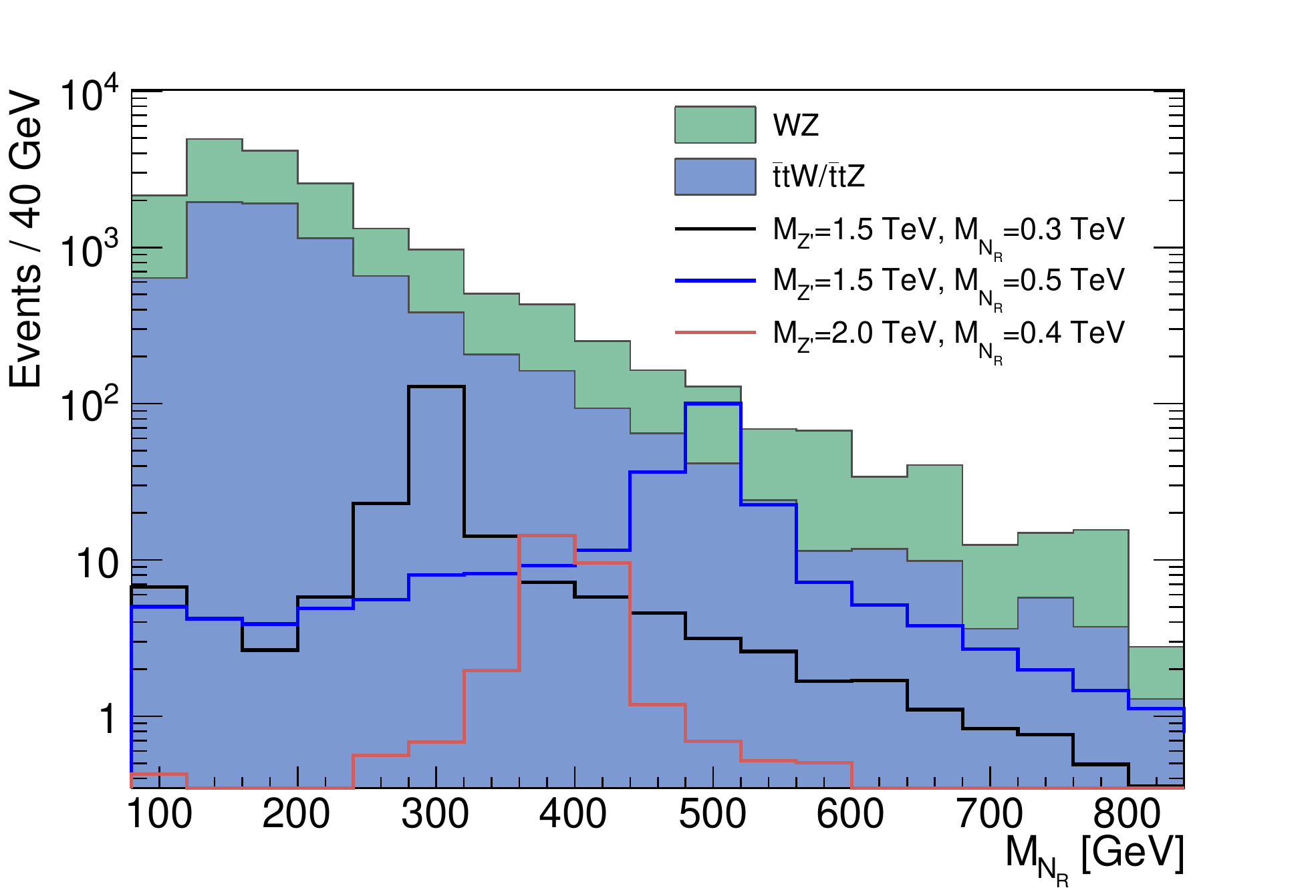} 
  \includegraphics[width=0.5\textwidth]{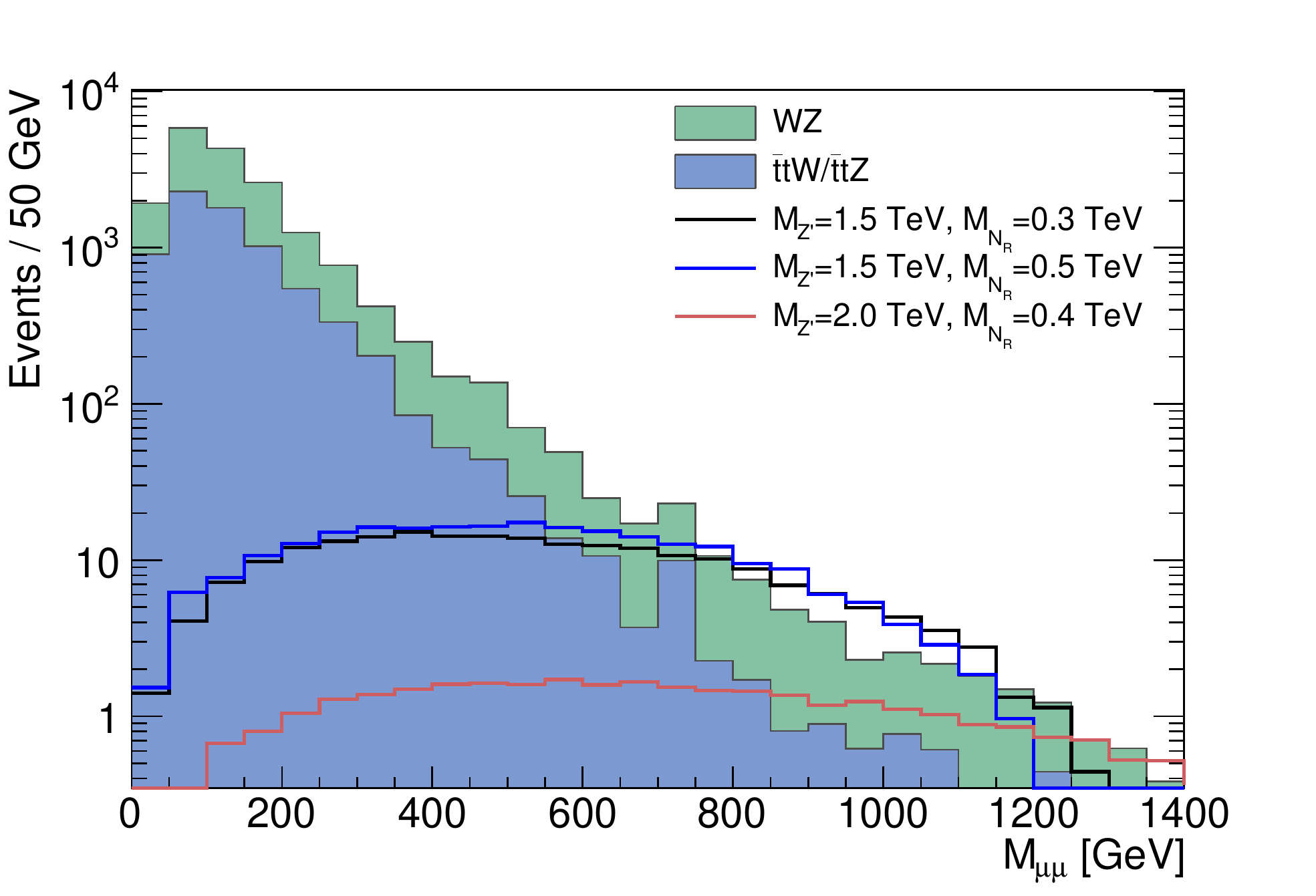} 
  \caption{Reconstructed $N_R$ mass (top) and same-sign di-muon invariant mass (bottom) after initial event selection, showing several benchmark signal points.}
  \label{fig:sig+bkg}
\end{figure}

After the initial selection detailed in Sec.~\ref{sec:initial_selection}, the SM background still dominates over any potential signal, as can be clearly seen in Fig.~\ref{fig:sig+bkg}. 
However, we have not yet exploited the fact that the same-sign muons produced by our signal are expected to have large $p_T$ and be well-separated.
The di-muon invariant mass can therefore be used to provide a good discriminator between the background and expected signal, as is the case in existing analyses (eg. \cite{1412.0237}). The same-sign di-muon invariant mass distribution is shown in Fig.~\ref{fig:sig+bkg}.

We have investigated the expected sensitivity of this search as a function of the selection cut on $M_{\mu\mu}$. 
We considered 15 benchmark signal points with $Z^\prime$ masses in the range 1--3 TeV and $N_R$ masses equal to $1/3$, $1/4$ and $1/5$ of the $Z^\prime$ mass. 
In order to obtain good sensitivity across a range of $Z^\prime$ and $N_R$ masses, we find that it is sufficient to define two (overlapping) signal regions: (i) $M_{\mu\mu}>500\,$GeV and (ii) $M_{\mu\mu}>800\,$GeV. 
These two regions provide optimal sensitivity for low and high $Z^\prime$ masses respectively, with the transition occurring between 1.5 -- 2 TeV. 
Lastly, we add a cut on the reconstructed RH neutrino mass following  Eq.~\ref{eq:masscut}. 
It is remarkable that this greatly increases the signal significance, as shown in  Tab.~\ref{tab:cutflow}.

\section{Results}

The expected number of signal and background events with $3000\,\text{fb}^{-1}$ at $\sqrt{s}=14\,$TeV are given in Tab.~\ref{tab:cutflow}, for a benchmark signal point. 
As expected, the same-sign di-muon invariant mass cut is highly effective at reducing the SM background. 
Furthermore, one can clearly see the improvement in sensitivity that can be obtained by reconstructing the RH neutrino mass. 

\begin{table}[ht]
  \begin{tabular}{| c | c c | c |}
    \hline
    & Background & Signal & $S/\sqrt{B}$ \\[-1.5ex]
    & & {\scriptsize $M_{Z^\prime}=2.0\,$TeV} & \\[-1.5ex] 
    & & {\scriptsize $M_{N_R}=0.5\,$TeV} & \\
    \hline
    Initial Selection & $1.79\times10^4$ & 35.8 & 0.3\\
    $M_{\mu\mu}>800\,$GeV & 32.5 & 14.8 & 2.6 \\
    $|M_{N_R}^\text{reco}-M_{N_R}|<0.1M_{N_R}$ & 5.0 & 10.5 & 4.7\\
    \hline
  \end{tabular}
  \caption{Expected number of events with $3000\,\text{fb}^{-1}$ for a benchmark signal point.}
  \label{tab:cutflow}
\end{table}

\begin{figure}[ht]
  \centering
  \includegraphics[width=0.5\textwidth]{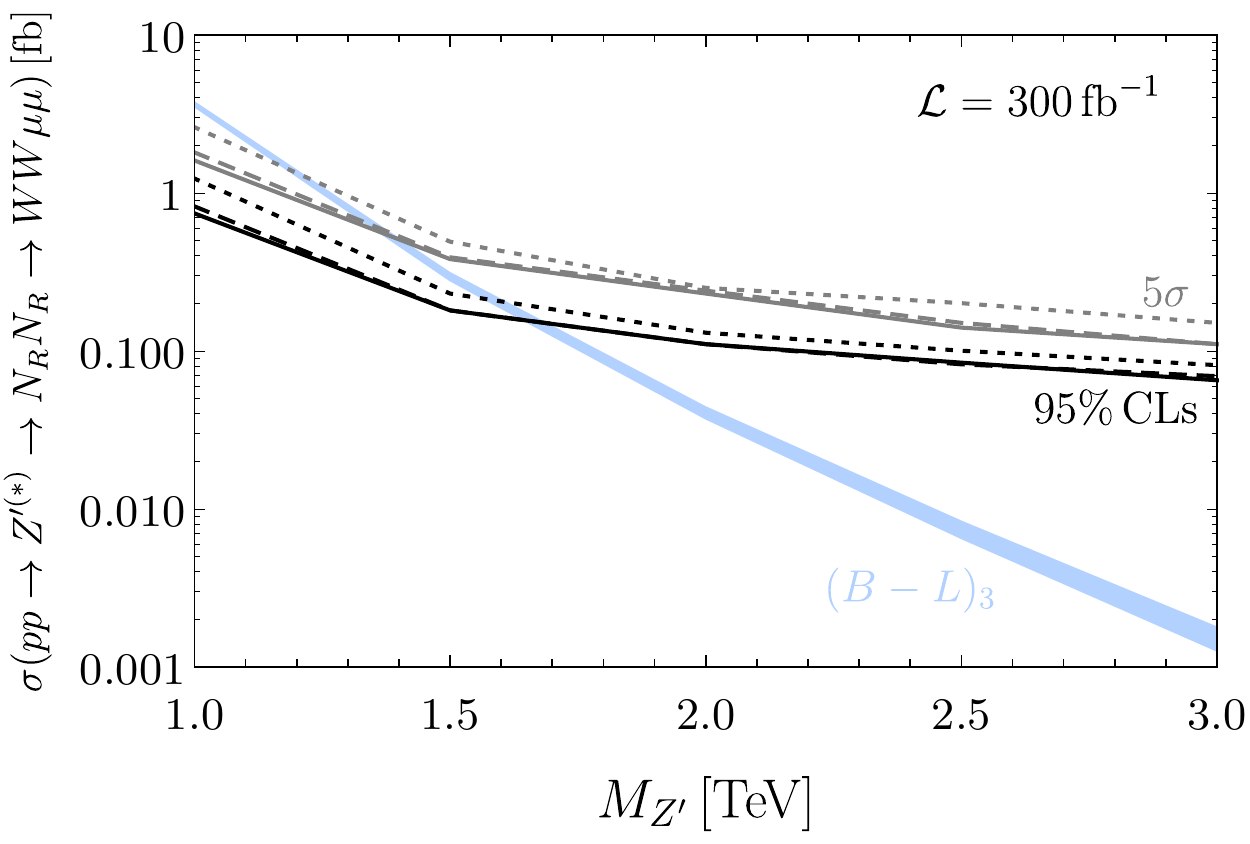} 
  \includegraphics[width=0.5\textwidth]{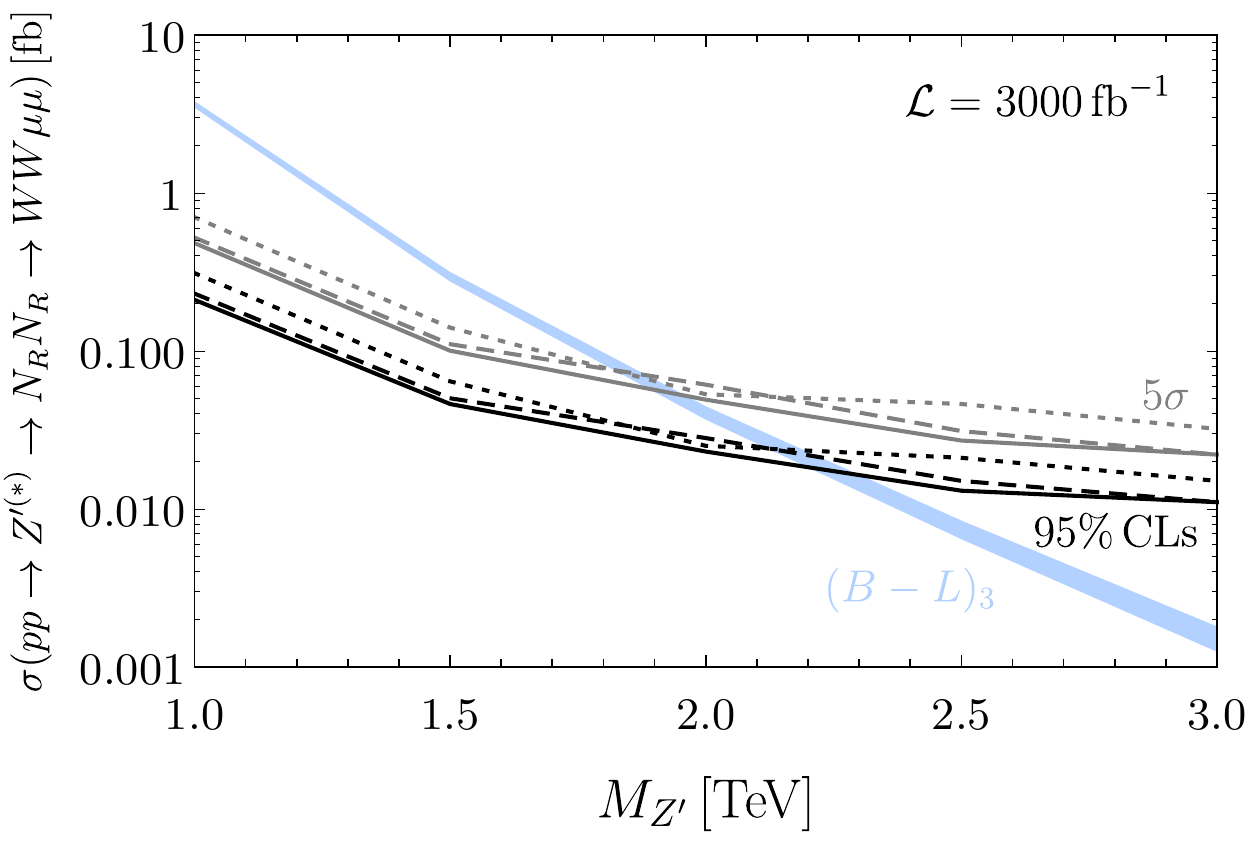} 
  \caption{Projected sensitivity with $300\,\text{fb}^{-1}$ (top) and $3000\,\text{fb}^{-1}$ (bottom) at $\sqrt{s}=14\,$TeV. The solid, dashed and dotted lines correspond to $M_{N_R}=(1/3,1/4,1/5)\times M_{Z^\prime}$ respectively. The blue band shows the expected cross-section and its associated uncertainty in the $(B-L)_3$ model, assuming $g=0.6$ and $M_{N_R}=M_{Z^\prime}/4$.}
  \label{fig:results}
\end{figure}

The main results of our analysis are given in Fig.~\ref{fig:results}, which shows the projected sensitivity with $300\,\text{fb}^{-1}$ and $3000\,\text{fb}^{-1}$ integrated luminosity at $\sqrt{s}=14\,$TeV. 
Results are presented in terms of $5\sigma$ discovery reach and 95\%~CLs exclusion limits on $\sigma(pp \to Z^\prime \to N_RN _R \to WW \mu\mu)$, as a function of the $Z^\prime$ mass. 
Limits are computed using {\tt RooStats}~\cite{1009.1003} and the asymptotic formulae for the profile likelihood~\cite{1007.1727}. 
We assume the following gaussian systematic uncertainties: background normalisation (10\%), signal efficiency (5\%) and luminosity (2.8\%). 
For comparison, we also show in Fig.~\ref{fig:results} the expected cross-section in the $U(1)_{(B-L)_{3}}$ model, with $g=0.6$ and $M_{N_R}=M_{Z^\prime}/4$. 
The cross-section is calculated at NLO in the 5-flavour scheme using {\tt MadGraph}. 
Note that there is a large uncertainty arising from the $b$ quark PDF.

We find that at the HL-LHC it will be possible to exclude $N_R$ production in the $U(1)_{(B-L)_{3}}$ model for $Z^\prime$ masses up to $\sim2.2\,$TeV. 
This allows one to probe RH neutrino masses in the range $0.2\lesssim M_{N_R} \lesssim 1.1\,$TeV.
Our results are relatively independent of the $N_R$ mass, provided that the $Z^\prime\to N_RN_R$ decay mode is kinematically allowed; however, for heavier $Z^\prime$ masses $N_R$ production via an off-shell $Z^\prime$ can contribute significantly to the total cross-section. 
Note that this reach is comparable to direct searches for the $Z^\prime$ in the $\tau\tau$ final state, assuming a naive extrapolation of the current limits. 
With $300\,\text{fb}^{-1}$ one can probe $N_R$ production for $Z^\prime$ masses up to $\sim1.7\,$TeV. 
In this case, the sensitivity at lower $Z^\prime$ masses ($\lesssim1\,$TeV) could be improved by introducing an additional signal region with a relaxed $M_{\mu\mu}$ cut.

\section{Conclusion}

We have investigated RH neutrino pair production in general $Z^\prime$ models at the (HL-)LHC. 
Focusing on final-states containing a pair of same-sign muons, we proposed a new, model-independent search. 
A novel aspect of this analysis is the ability to reconstruct the RH neutrino, with good mass resolution.
This leads to a significantly improved sensitivity over general same-sign di-lepton searches. 
Within the $U(1)_{(B-L)_{3}}$ model, we find that in the future it will be possible to probe RH neutrino masses in the range $0.2\lesssim M_{N_R} \lesssim 1.1\,$TeV, or equivalently $Z^\prime$ masses up to $\sim2.2\,$TeV.
While we focused on the di-muon channel due to reduced backgrounds, a similar analysis could be performed for $e^\pm e^\pm$ and $e^\pm \mu^\pm$ final states. 
Finally, our analysis could also be straightforwardly adapted to RH neutrino production from a $W^\prime$ boson, utilising the same techniques to reconstruct the $N_R$ mass.

\vspace{1ex}

\noindent {\bf{Acknowledgements}}
This work is supported by Grants-in-Aid for Scientific Research from the Ministry of Education, Culture, Sports, Science, and Technology (MEXT), Japan, No. 26104009 (T.T.Y.), No. 16H02176 (T.T.Y.) and No. 17H02878 (T.T.Y.), and by the World Premier International Research Center Initiative (WPI), MEXT, Japan (P.C., C.H,. and T.T.Y.). 

\newpage

\bibliography{NR-LHC}

\end{document}